\documentclass[12pt,english,floatfix,nofootinbib,superscriptaddress,aps,prd,preprint]{revtex4}
\usepackage[utf8]{inputenc}
\usepackage{float}
\usepackage{array}

\usepackage{verbatim}
\usepackage{xcolor} 
\usepackage{lipsum}
\usepackage{graphicx}
\usepackage{amsmath,amsthm,amsfonts,amssymb}
\usepackage{graphicx}
\usepackage[english]{babel}
\usepackage{color}

\usepackage{tensor}
\usepackage{esint}
\usepackage[dvips]{epsfig}
\usepackage[dvips]{graphicx}
\usepackage{float}
\usepackage{units}
\usepackage{textcomp}
\usepackage{mathrsfs}
\usepackage{amsmath}
\usepackage[makeroom]{cancel}
\usepackage{amssymb}
\usepackage{amsbsy}
\usepackage{amsfonts}
\usepackage{amssymb,mathrsfs,xcolor}
\usepackage{esint}
\usepackage{array}
\usepackage{graphicx}
\usepackage{mathtools,slashed}
\usepackage{slashed}
\usepackage{natbib}
 \usepackage{booktabs}
 
\usepackage{hyperref}

\hypersetup{
    colorlinks,
    citecolor=blue,
    filecolor=green,
    linkcolor=purple,
    urlcolor=red,
}

\newcommand{\ie}{\begin{equation}}
\newcommand{\fe}{\end{equation}}
\newcommand{\se}{\begin{eqnarray}}
\newcommand{\ff}{\end{eqnarray}}

\usepackage{hyperref}
\hypersetup{colorlinks,breaklinks,
			citecolor=[rgb]{0,0.0,1.0},
            urlcolor=[rgb]{0.0,0.0,1.0},
            linkcolor=[rgb]{0,0.5,0.9}}

\begin{document}

\title{Rotating black hole shadows in metric-affine bumblebee gravity}

\author{Jose R. Nascimento}
\email[\textit{The authors are listed in alphabetical order.}\protect\\
Electronic address: ]{jroberto@fisica.ufpb.br}
\affiliation{Departamento de Física, Universidade Federal da Paraíba, Caixa Postal 5008, 58051-970, João Pessoa, Paraíba,  Brazil}

\author{Ana R. M. Oliveira}
\email{ana.rafaely@academico.ufpb.br}
\affiliation{Departamento de Física, Universidade Federal da Paraíba, Caixa Postal 5008, 58051-970, João Pessoa, Paraíba,  Brazil} 

\author{Albert Yu. Petrov}
\email{petrov@fisica.ufpb.br}
\affiliation{Departamento de Física, Universidade Federal da Paraíba, Caixa Postal 5008, 58051-970, João Pessoa, Paraíba,  Brazil} 

\author{Paulo J. Porfírio}
\email{pporfirio@fisica.ufpb.br}
\affiliation{Departamento de Física, Universidade Federal da Paraíba, Caixa Postal 5008, 58051-970, João Pessoa, Paraíba,  Brazil} 

\author{Amilcar R. Queiroz}
\email{amilcarq@df.ufcg.edu.br}
\affiliation{Departamento de Física, Universidade Federal de Campina Grande, Caixa Postal 10071, 58429-900, Campina Grande, Paraíba, Brazil}


\date{\today}

\begin{abstract}
  In this work, we investigate the structure of black hole shadows in the bumblebee gravity model formulated within the metric-affine framework, which incorporates spontaneous Lorentz symmetry breaking (LSB) through a vector field $B_\mu$ with a non-zero vacuum expectation value. We analyze the influence of the dimensionless rotation parameter $a = J/M$ and the  Lorentz-violating (LV) coefficient $X = \xi b^2$ on the photon sphere radius, the critical impact parameter, and the shadow morphology. Using ray-tracing simulations with the GYOTO code and accretion disks, we observe that increasing values of $X$ induce progressive vertical flattening, asymmetric ``teardrop''-shaped deformations, and local collapse of the lower silhouette region, interacting with the rotational Doppler effect. These anisotropic signatures distinguish the bumblebee model from the standard Kerr metric and provide observational tests for  LSB effects in strong gravity regimes, potentially detectable by the Event Horizon Telescope in sources such as M87* and Sgr A*.
\end{abstract}

\maketitle

\section{Introduction}

Black holes are one of the most intriguing solutions of general relativity (GR), they provide a natural setting for testing gravity in the strong field regime. Recent observations by the Event Horizon Telescope (EHT), including the imaging of M87* \cite{Akiyama2019,Akiyama2019b,Akiyama2019c,Akiyama2019d,Akiyama2019e,Akiyama2019f,Akiyama2019a,akiyama2019first} and Sgr A* \cite{EHT2022}, have opened a new avenue for investigating the near-horizon region of these compact objects. 

A striking physical phenomenon is gravitational lensing, that is the deflection of light by massive objects. In the strong field regime, light rays passing close to a compact object may undergo large deflections (including turning around the gravitational lens multiple times) and several methods have been developed to describe this behavior, including the approaches introduced by Tsukamoto \cite{Tsukamoto:2016jzh} and Bozza \cite{Bozza:2001xd, Bozza:2002zj}. These formalisms have been widely applied to the study of light propagation in curved spacetimes \cite{Nascimento:2020ime, Filho:2024zxx, Filho:2024isd, Soares:2025hpy, Nakajima:2012pu, Tsukamoto:2012xs, Tsukamoto:2016qro, Chen:2016hil, Bozza:2002af, Bozza:2005tg, Bozza:2006nm, Soares:2024rhp, Virbhadra:2024xpk}. In particular, black hole shadows \cite{Perlick:2021aok} may be regarded as strong field lensing phenomena, since their boundaries are determined by light rays asymptotically approaching unstable photon orbits and therefore appear as dark regions in the observer’s sky. Such effects can be useful to probe possible departures from GR.

Among several models that extend GR, theories that incorporates spontaneous LSB constitute one of the most paradigmatic classes \cite{judes2003,robertson1949,myers2003,bertolami2005,reyes2008,mattingly2008,rubtsov2014,liberati2013,tasson2014,hees2016,rovelli2004,kostelecky1989a,kostelecky1989b}. A particular example is the bumblebee gravity, in which a vector field $B_\mu$, acquires a non-zero vacuum expectation value, $\langle B_{\mu}\rangle=b_{\mu}$, and thereby selects a preferred direction in spacetime. Within the metric-affine (Palatini) approach, different formulations of bumblebee gravity have been developed \cite{Delhom2021,Delhom2022,Delhom2022a,Lambiase2023,Ara_jo_Filho_2024,AraujoFilho2023}. In particular, it was obtained black hole solutions that can be viewed as LSB generalizations of the Schwarzschild and Kerr geometries in the traceless version \cite{Ara_jo_Filho_2024,AraujoFilho2023}.  It is worth mentioning that both versions present projective invariance, which means the avoidance of undesirable ghost degrees of freedom \cite{kostelecky1989c,kostelecky1991,kostelecky1995,gambini1999,bojowald2005,amelino2000,carroll2001,modesto2012,nascimento2021,klinkhamer2004,bernadotte2007,klinkhamer1998,klinkhamer2000,klinkhamer2002,ghosh}.

In this work, we perform a systematic analysis of the shadow of the stationary axisymmetric black hole solution of the metric-affine traceless bumblebee gravity Ref.~\cite{Ara_jo_Filho_2024}.  Our main goal is to investigate how the effects of the rotation parameter $a$ and the  LSB parameter $X$ affect the photon sphere,  the critical impact parameter, and the resulting shadow morphology.  We complement the analytical discussion with numerical intensity profiles and ray-tracing simulations performed using \textsc{GYOTO}, including a thin accretion-disk model and different observational configurations. Our results show that the LSB sector can induce visible departures from the standard Kerr shadow, including vertical flattening, asymmetric deformations, and teardrop-like structures, whose impact increases when LSB and rotation act simultaneously.



 The paper is organized as follows. In Section~2, we provide a brief review of the bumblebee metric. In Section~3, we discuss the construction of the shadow and its dependence on the parameters $a$ and $X$. In Section~4,  we analyze the shadow through the corresponding intensity profile. In Section~5, we highlight the relativistic effects on shadows using \texttt{GYOTO}. 
 In Section~6, we summarize our results and discuss their implications.

\section{Rotating black hole solution in metric-affine traceless bumblebee gravity} 
\label{sec:traceless-mabg}

In this section, we briefly review the metric-affine (\emph{Palatini}) formulation of the bumblebee model in its traceless version.  As it is well known, the metric $g_{\mu\nu}$ and the connection $\Gamma^\lambda_{\ \mu\nu}$ are treated as independent variables in the Palatini approach. This model is characterized by a vector field $B_\mu$  which acquires a nontrivial vacuum expectation value $\langle B_\mu \rangle = b_\mu$, generating  LSB. The action of the model is defined as \cite{Ara_jo_Filho_2024,AraujoFilho2023}
\begin{equation}
 \begin{aligned}\label{action}
\mathcal{S}_{B}&=\int d^4x\sqrt{-g}\left[\frac{1}{2\kappa^2}\left(R(\Gamma)+\xi\left(B^\mu B^\nu-\frac{1}{4}B^2g^{\mu\nu}\right)R_{\mu\nu}(\Gamma)\right)-\frac{1}{4}B^{\mu\nu}B_{\mu\nu}-\right.\\&-\left.V(B^{\mu}B_{\mu}\pm b^{2})\right]+\int d^{4}x\sqrt{-g}\mathcal{L}_{mat}(g_{\mu\nu},\psi),
\end{aligned}   
\end{equation}

\noindent where $B_{\mu\nu} \equiv \partial_\mu B_\nu - \partial_\nu B_\mu$,  $b^2 \equiv g^{\mu\nu} b_\mu b_\nu$, $\xi$ is the non-minimal coupling and $R(\Gamma) \equiv g^{\mu\nu} R_{\mu\nu}(\Gamma)$. The  requirement for the potential $V$ to yield a minimal value  enforces $\langle B_\mu \rangle = b_\mu$ while keeping $b^2$ constant.\footnote{In this work, we assume that $\mathcal{L}_{\rm mat}$ does not couple directly to the independent connection.}
The action \eqref{action} is equivalent to a subclass of the gravitational sector of the SME with LSB coefficients $(u, s_{\mu\nu}, t_{\mu\nu\alpha\beta})$ such that

\begin{equation}
u=0,\qquad s_{\mu\nu}=\xi\Big(B_\mu B_\nu-\tfrac{1}{4}B^2 g_{\mu\nu}\Big),\qquad t_{\mu\nu\alpha\beta}=0,
\label{eq:sme1}
\end{equation}

\noindent or equivalently, after absorbing the trace part of $s_{\mu\nu}$ into $u$,

\begin{equation}
u=\frac{\xi}{4}B^2,\qquad s_{\mu\nu}=\xi B_\mu B_\nu,\qquad t_{\mu\nu\alpha\beta}=0.
\label{eq:sme2}
\end{equation}

The model is invariant under projective transformations of the connection:

\begin{equation}
\Gamma^\mu_{\ \nu\alpha}\rightarrow \Gamma^\mu_{\ \nu\alpha}+\delta^\mu_\alpha A_\nu,
\label{eq:projective}
\end{equation}

\noindent which implies that the symmetric part of the Ricci tensor remains invariant. This invariance, typical of Ricci-based gravities \cite{Afonso:2017bxr, BeltranJimenez:2017doy} relying on $R_{\mu\nu}$, prevents ghost degrees of freedom in the gravitational sector.

Varying \eqref{action} with respect to $g_{\mu\nu}$ yields a modified Einstein equation, involving the  matter energy-momentum tensor $T^{\rm mat}_{\mu\nu}$ and that of the bumblebee sector,

\begin{equation}
T^{B}_{\mu\nu}=B_{\mu\alpha}B^{\ \alpha}_\nu-\frac{1}{4}g_{\mu\nu}B_{\alpha\beta}B^{\alpha\beta}-V\,g_{\mu\nu}+2V'(B^2\!\pm b^2)\,B_\mu B_\nu.
\end{equation}

Through suitable contractions (with $g^{\mu\nu}$ and powers of $B^\mu$), one can solve  the equations of motion for $R(\Gamma)$, $B^\mu R_{\mu\nu}(\Gamma)$, and $B^\mu B^\nu R_{\mu\nu}(\Gamma)$, and recast the dynamics as an effective Einstein equation for  auxiliary metric $h_{\mu\nu}$, which is disformally related with $g_{\mu\nu}$ (see \cite{Ara_jo_Filho_2024,AraujoFilho2023} for the detailed derivation of the field equations):

\begin{equation}
\begin{aligned}R_{\mu\nu}(h)&=\kappa_{eff}^2\bigg\{T_{\mu\nu}-\frac{1}{2}g_{\mu\nu}T+\frac{2\xi g_{\mu\nu}}{(4+5\xi B^2)}\bigg[B^{\alpha}B^{\beta}T_{\alpha\beta}-\frac{B^2T}{16}(4-3\xi B^2)\bigg]+\\&+\frac{8\xi}{(4+3\xi B^{2})}B_{(\mu}\left[T_{\nu)\alpha}B^{\alpha}-\frac{B_{\nu)}T}{2}-\right.\\&- \frac{2\xi B_{\nu)}}{(4+5\xi B^{2})}\left(B^{\alpha}B^{\beta}T_{\alpha\beta}-\frac{1}{4}B^{2}T\left(1-\frac{3}{4}\xi B^{2}\right)\right)\bigg] \bigg\},\end{aligned}
\label{eq:hEinstein}
\end{equation}
with
\begin{equation}
    \kappa^2_{\rm eff}\equiv\frac{\kappa^2}{1-\frac{\xi}{4}B^2}.
\end{equation}

This shows that the entire geometric dynamics can be organized in the $h$ frame, leaving $g_{\mu\nu}$ aside. In this context, we study a stationary axisymmetric solution in the traceless metric-affine bumblebee model, extending the analysis through the intensity profile and examining its relativistic effects.

The bumblebee field satisfies the following field equation (obtained by varying the action \eqref{action} with respect to $B_{\mu}$)
\begin{eqnarray}
\nabla^{(g)}_{\mu}B^{\mu\alpha}=\mathcal{M}^{\alpha}_{\,\,\,\nu}B^{\nu},
\label{proca}
\end{eqnarray}
where the effective mass-squared tensor has been defined by
\begin{eqnarray}
\nonumber \mathcal{M}^{\alpha}_{\,\,\,\nu}&=&\bigg\{2V^{\prime}+\frac{\xi T\left(4-3\xi B^2\right)}{4\left(4+3\xi B^2\right)}+\frac{8\xi^2}{\left(4+3\xi B^2\right)\left(4+5\xi B^2\right)}\bigg[B^{\mu}B^{\lambda}T_{\mu\lambda}-\\
&-&\frac{1}{4}B^2 T\left(1-\frac{3}{4}\xi B^2 \right)\bigg]\bigg\}\delta^{\alpha}_{\,\,\,\nu}-\frac{4\xi}{\left(4+3\xi B^2\right)}T^{\alpha}_{\,\,\,\nu},
\label{Mat}
\end{eqnarray}
where the prime indicates the derivative with respect to the argument of the potential $V$, and $\nabla^{(g)}_{\mu}$ is the covariant derivative related to the Levi-Civita connection of the metric $g_{\mu\nu}.$

\section{Construction of Shadows}

In Ref.~\cite{Ara_jo_Filho_2024}, the authors investigate  LSB through the metric-affine traceless bumblebee model metric-affine generalization of the gravitational sector of the SME, incorporating the  LSB coefficients $u$ and $s_{\mu\nu}$, which, in particular, yield a metric-affine generalization of the gravitational sector.

In this context, we compute the black hole shadow using the exact stationary and axisymmetric vacuum solution, specifically the physical metric and the bumblebee profile derived from the field equations \eqref{eq:hEinstein} and \eqref{proca}. Accordingly, the line element is given by {  \cite{Ara_jo_Filho_2024}:}

\begin{equation}\label{linha_element}
\begin{aligned}
\mathrm{d}s_{(g)}^2&=-\left(\frac{\Delta-a^{2}\sin^{2}\theta}{\rho^{2}}\right)\frac{\mathrm{d}t^{2}}{\sqrt{\left(1+\frac{3X}{4}\right)\left(1-\frac{X}{4}\right)}}-\frac{4aMr\sin^{2}\theta}{\sqrt{\left(1+\frac{3X}{4}\right)\left(1-\frac{X}{4}\right)\rho^{2}}}\mathrm{d}t\mathrm{d}\phi+\\&+\frac{1}{\Delta\sqrt{\left(1+\frac{3X}{4}\right)\left(1-\frac{X}{4}\right)}}\left(a^{2}\cos^{2}\theta+r^{2}\frac{\left(1+\frac{3X}{4}\right)}{\left(1-\frac{X}{4}\right)}\right)\mathrm{d}r^{2}+\\&+\frac{1}{\sqrt{\left(1+\frac{3X}{4}\right)\left(1-\frac{X}{4}\right)}}\left(r^{2}+a^{2}\cos^{2}\theta\frac{\left(1+\frac{3X}{4}\right)}{\left(1-\frac{X}{4}\right)}\right)\mathrm{d}\theta^{2}+\\&+\frac{(r^2+a^2)^2-a^2\Delta\sin^2\theta}{\sqrt{\left(1+\frac{3X}{4}\right)\left(1-\frac{X}{4}\right)\rho^2}}\sin^2\theta\mathrm{d}\phi^2+\\&+\frac{2rXa\cos\theta}{\sqrt{\left(1+\frac{3X}{4}\right)}\left(1-\frac{X}{4}\right)^{\frac{3}{2}}}\frac{\mathrm{d}r\mathrm{d}\theta}{\sqrt{\Delta}},&
\end{aligned}
\end{equation}

\noindent  where $\Delta = r^2 - 2Mr + a^2$, $\rho^2 = r^2 + a^2 \cos^2\theta$, and $X = \xi b^2$ is the dimensionless LSB parameter, with $\xi$ being the non-minimal coupling constant and $b^2 = b_\mu b^\mu$ the squared vacuum expectation value of the bumblebee field $B_\mu$.

By definition, the black hole shadow represents the apparent boundary in the sky of a distant observer, delineating the directions from which light rays are captured by the event horizon versus those that escape to infinity. This silhouette is shaped by unstable null geodesics, particularly the photon sphere.

\subsection*{A. Analytical Structure for Null Geodesics}

The calculation of the shadow begins with the analysis of null geodesics in spacetime. The Lagrangian describing the geodesic motion is:
\begin{equation}
    \mathcal{L} = g_{\mu\nu} \dot{x}^\mu \dot{x}^\nu = 0 \,
\end{equation}
for null curves, where the dot denotes differentiation with respect to an affine parameter $\lambda$. Due to axisymmetry, there are two conserved quantities: the energy $E = -g_{t\mu} \dot{x}^\mu$ and the angular momentum $L = g_{\phi\mu} \dot{x}^\mu$. Restricting to the equatorial plane ($\theta = \pi/2$) for simplicity in the initial analysis,  one can write the radial equation for null geodesics  as

\begin{equation}
\dot{r}^2 = \left(1+\frac{3X}{4}\right)^{1/2} \left(1-\frac{X}{4}\right)^{3/2} \frac{1}{r^4} (E - V^+)(V^- - E),
\end{equation}

where the effective potentials are
\begin{equation}\label{potencial}
\begin{aligned}
V_{\pm} = & \frac{1}{a^2(2M + r) + r^3} (\pm r \sqrt{-((X - 4)(3X + 4))})  \times \left[ \sqrt{(a^2 + r(r - 2M))} \right. \\
& \times \sqrt{-\frac{4\mathcal{L}\sqrt{-((X - 4)(3X + 4))}(a^2(2M + r) + r^3) + L^2r(X - 4)(3X + 4)}{r(X - 4)^2(3X + 4)^2}} \\
& \left. + 2aLM \right].
\end{aligned}
\end{equation}






{ The photon sphere radius is obtained by simultaneously imposing $\dot{r} = 0$ and 
$\ddot{r} = 0$ on the radial null geodesic equation. As shown in the previous section, 
in the equatorial plane ($\theta = \pi/2$), the off-diagonal $dr\,d\theta$ term of the 
metric~\eqref{linha_element} vanishes identically, since it carries a factor proportional to $\cos\theta$. 
Furthermore, note that for circular orbits ($dr=0$), all remaining metric components share the same overall factor $\alpha^{-1}$, 
where
\begin{equation}\label{eq:alpha}
    \alpha \equiv \sqrt{\left(1+\frac{3X}{4}\right)\left(1-\frac{X}{4}\right)} = \text{const},
\end{equation}
so that the metric of the equatorial $(t, \phi)$ sector can be cast into the form $g_{\mu\nu}\Big|_{\theta=\pi/2,\, r=const.} = \tilde{g}_{\mu\nu}/\alpha$, with $\tilde{g}_{\mu\nu}$ 
denoting the standard Kerr components. This common factor cancels in both the 
circular-orbit conditions $N(r_{ph}) = 0$ and $N'(r_{ph}) = 0$, where
\begin{equation}
    N(r) \equiv -E^2\tilde{g}_{\phi\phi} - 2EL\tilde{g}_{t\phi} - L^2\tilde{g}_{tt},
\end{equation}
rendering the photon sphere equation identical to that of the Kerr metric. 
Consequently, the photon sphere radius is {independent of the LSB parameter} 
$X$ and is given by the standard Kerr result:
\begin{equation}
    r_{ph}^{(\pm)} = 2M\left[1 + \cos\!\left(\frac{2}{3}\arccos\!\left(\mp\frac{a}{M}
    \right)\right)\right],
    \label{eq:rph}
\end{equation}
where the upper (lower) sign corresponds to the retrograde (prograde) photon orbit. 
In the limit $a \to 0$, one recovers $r_{ph} = 3M$, as expected for the Schwarzschild 
spacetime.
 
}

%
 
%
%

\section{Analyzing the shadow with the intensity profile}\label{sec:shadow}

In this section, we will analyze in more detail the shadows for the case of the bumblebee model, to identify signatures that depend on the LSB parameter $X$ and the rotation $a$, respectively. By definition, the shadow is defined by the instability of the photon orbit, and the radius of this orbit is identified when we consider the observer at infinity. { As remarked in Section III, the LSB rotating metric reduces to the Kerr metric multiplied by a constant overall factor. In this case,  the two critical 
impact parameters associated with the prograde and retrograde photon orbits coincide with those of the Kerr metric, namely,
\begin{equation}\label{eq:bc}
    b_c^{(\pm)} = \frac{-\tilde{g}_{t\phi}(r_{ph}^{(\pm)}) \pm 
    \sqrt{\tilde{g}_{t\phi}^2(r_{ph}^{(\pm)}) - \tilde{g}_{tt}(r_{ph}^{(\pm)})\,
    \tilde{g}_{\phi\phi}(r_{ph}^{(\pm)})}}{-\tilde{g}_{tt}(r_{ph}^{(\pm)})}.
\end{equation}
{ To obtain the explicit forms of $b_c^{(\pm)}$, we first establish the general identity, valid for 
any $r$ in the equatorial plane of the Kerr geometry, initiated in the equation \eqref{eq:bc}, continued in the equation below

\begin{equation}
\tilde{g}^{2}_{t\phi}(r) - \tilde{g}_{tt}(r)\,\tilde{g}_{\phi\phi}(r)
\Big|_{\theta=\pi/2}
=\frac{4a^{2}M^{2}}{r^{2}}
+\frac{(r-2M)}{r}\!\left(r^{2}+a^{2}+\frac{2Ma^{2}}{r}\right)
= r^{2}-2Mr+a^{2} \equiv \Delta(r),
\label{eq:discriminant}
\end{equation}
\noindent where $\Delta(r)=r^{2}-2Mr+a^{2}$.
Performing the same procedure for $r_{ph}$ now, we find the same

\begin{equation}
\tilde{g}^{2}_{t\phi}-\tilde{g}_{tt}\tilde{g}_{\phi\phi}
\bigg|_{r^{(\pm)}_{\rm ph}}
= 4M^{2}
\cos\!\left(\frac{2}{3}\arccos\!\left(\mp\frac{a}{M}\right)\right)
\left[1+\cos\!\left(\frac{2}{3}\arccos\!\left(\mp\frac{a}{M}
\right)\right)\right]+a^{2}.
\label{eq:sqrt_delta_rph}
\end{equation}
Substituting equations \eqref{eq:rph} and \eqref{eq:sqrt_delta_rph} into equation~\eqref{eq:bc}, one finds an exact expression for the critical impact parameters 

\begin{equation}
\begin{split}
b^{(\pm)}_c = \frac{1}{\cos\!\left(\dfrac{2}{3}\arccos\!\left(\mp\dfrac{a}{M}\right)\right)}
\bigg\{
a \pm \left[1+\cos\!\left(\frac{2}{3}\arccos\!\left(\mp\frac{a}{M}\right)\right)\right]
\\
\times\sqrt{4M^{2}\cos\!\left(\frac{2}{3}\arccos\!\left(\mp\frac{a}{M}
\right)\right)\left[1+\cos\!\left(\frac{2}{3}\arccos\!\left(\mp
\frac{a}{M}\right)\right)\right]+a^{2}}\,
\bigg\},
\end{split}
\label{eq:bc_full_explicit}
\end{equation}
 This is manifestly independent of $X$ and depends only on $M$ 
and $a$.


Note that the analytical expressions for critical parameters are fully consistent with the numerical values $b_{\rm crit} = 3\sqrt{3}\,M \approx 5.1962\,M$ reported 
in Tables~\ref{tab:II} and~\ref{tab:III} for all values of $X$ and $a$.
}{ While the critical impact parameter $b_c$ in the equatorial plane is 
independent of $X$, the full 
shadow boundary is a two-dimensional curve in the observer's sky, and its morphology is sensitive to the LSB parameter whenever $a \neq 0$. 
To better understand this, we recall that the metric \eqref{linha_element} contains 
an off-diagonal term
\begin{equation}\label{eq:g_rteta}
g_{r\theta} = \frac{2rXa\cos\theta}
{\left[\left(1+\tfrac{3X}{4}\right)
\left(1-\tfrac{X}{4}\right)\right]^{3/2}\sqrt{\Delta}},
\end{equation}
which carries a factor $\alpha^{-3/2}$ rather than $\alpha^{-1}$. 
Consequently, off-equatorial metric components do not share the 
same global conformal factor as the equatorial ones, and the cancellation 
argument no longer applies. This term vanishes identically for $a = 0$ 
or $X = 0$, which immediately implies that any LSB-induced deformation 
of the shadow requires the simultaneous presence of both rotation and 
LSB.

\subsubsection*{Two-parameter description of the shadow boundary}

We will employ the Hamilton-Jacobi approach \cite{Carter, Chandrasekhar, Cunha} to describe the shadow boundary. The starting point is to write down the Hamilton-Jacobi equation for null geodesics, namely,
\begin{equation}
  H=\frac{1}{2}g^{\mu\nu}\partial_\mu S\,\partial_\nu S = 0,  
\end{equation}
with $S = -Et + L_z\varphi + W(r,\theta)$, the expansion yields
\begin{equation}
\underbrace{g^{tt}E^2 + 2g^{t\varphi}(-E)L_z
+ g^{\varphi\varphi}L_z^2}_{\text{azimuthal sector}}
+ \underbrace{g^{rr}p_r^2
+ g^{\theta\theta}p_\theta^2}_{\text{separates in Kerr}}
+ \underbrace{2g^{r\theta}p_r\,p_\theta}_{\text{breaks separability}}
= 0,
\end{equation}
where the canonical momenta are identified by $p_r=\partial S/\partial r$ and $p_\theta=\partial S/\partial\theta$.
For the standard Kerr metric, $g^{r\theta}=0$ and the equation separates
exactly, yielding the Carter constant $\mathcal{K}$ (see details in \cite{Carter}). In our case, the
non-vanishing $g^{r\theta}$ spoils this separation. It is worth stressing out that the presence of the mixed component $g_{r\theta}$ is a coordinate-dependent indication that the usual Kerr separability is no longer manifest in these coordinates. The coordinate-independent criterion, which is a stronger criterion, is the existence of a nontrivial rank-two Killing--Stäckel tensor, or a conformal Killing--Stäckel tensor for null geodesics, generating a Carter-like constant. Therefore, in what follows, the perturbative construction should be understood as a test of how the Kerr hidden symmetry is deformed by the LSB parameter $X$. Let us follow this methodology
 to obtain the first-order LSB correction to the Carter constant for the Kerr metric. For this purpose, we decompose the full inverse metric as
$g^{\mu\nu} = \tilde{g}^{\mu\nu}_{\rm Kerr} + X k^{\mu\nu} + \mathcal{O}(X^2)$,
where the first-order perturbation $k^{\mu\nu} =
-\tilde{g}^{\mu\alpha}k_{\alpha\beta}\tilde{g}^{\beta\nu}$ has components
\begin{equation}
\begin{aligned}
k^{tt} &= \tfrac{1}{4}\tilde{g}^{tt}, \qquad
k^{t\varphi} = \tfrac{1}{4}\tilde{g}^{t\varphi}, \qquad
k^{\varphi\varphi} = \tfrac{1}{4}\tilde{g}^{\varphi\varphi}, \\[6pt]
k^{rr} &= -\frac{\Delta\!\left(3r^2 - a^2\cos^2\theta\right)}{4\rho^4},
\qquad
k^{\theta\theta} = -\frac{3a^2\cos^2\theta - r^2}{4\rho^4},
\qquad
k^{r\theta} = -\frac{ar\cos\theta}{\sqrt{\Delta}\,\rho^4}.
\end{aligned}
\label{eq:kmunu}
\end{equation}
The perturbed null Hamiltonian is $H = H_0 + XH_1 + \mathcal{O}(X^2)$,
with $H_1 = \tfrac{1}{2}k^{\mu\nu}p_\mu p_\nu$. Substituting
equation~\eqref{eq:kmunu} and using the Kerr on-shell condition for null geodesics
$H_0 = 0$, which implies
\begin{equation}
\tilde{g}^{tt}E^2 - 2\tilde{g}^{t\varphi}EL_z + \tilde{g}^{\varphi\varphi}L_z^2
= -\frac{\Delta}{\rho^2}p_r^2 - \frac{1}{\rho^2}p_\theta^2,
\end{equation}
all diagonal contributions combine algebraically, and $H_1$ collapses to
the compact form
\begin{equation}
H_1 = -\frac{A^2}{2\rho^4}, \qquad
A \equiv r\sqrt{\Delta}\,p_r + a\cos\theta\,p_\theta.
\label{eq:H1compact}
\end{equation}
We seek a corrected invariant of the form
\begin{equation}
\mathcal{K} = \mathcal{K}_0 + X\mathcal{K}_1 + \mathcal{O}(X^2),
\label{eq:Kexpansion}
\end{equation}
where $\mathcal{K}_0$ is the standard Kerr Carter constant, defined as the conserved quantity associated with the separability of the
Hamilton-Jacobi equation in Kerr spacetime:
\begin{equation}
\mathcal{K}_0
= p_\theta^2
+ \cos^2\!\theta\!\left(\frac{L_z^2}{\sin^2\theta} - a^2 E^2\right).
\label{eq:K0def}
\end{equation}

The quantity $\mathcal{K}_0$ is exactly conserved in the Kerr geometry, namely $\{\mathcal{K}_0,H_0\}=0$. In the bumblebee-deformed geometry, however, the same Kerr expression is not assumed to be an exact invariant of the full Hamiltonian. Instead, we look for a deformed Carter-like quantity of the form $\mathcal{K}=\mathcal{K}_0+X\mathcal{K}_1+\mathcal{O}(X^2)$.
The observer's two impact parameters are defined as
\begin{equation}
\ell \equiv \frac{L_z}{E}, \qquad \eta \equiv \frac{\mathcal{K}_0}{E^2},
\label{eq:impact}
\end{equation}
which parametrize the shadow boundary as the photon reaches $r \to \infty$:
$\ell$ is the azimuthal impact parameter (angular momentum per unit energy)
and $\eta$ is the Carter parameter (quadratic invariant per unit energy squared).
Together they determine the apparent position
$(\alpha_{\rm obs}, \beta_{\rm obs})$ in the observer's sky.
 
The first-order correction $\mathcal{K}_1$ is the quantity that must
be determined: it encodes how the bumblebee LSB deformation modifies the conserved structure of the geodesic motion. Substituting the
expansion~\eqref{eq:Kexpansion} into the conservation condition
$\{\mathcal{K}, H\} = \mathcal{O}(X^2)$, one finds that $\mathcal{K}_1$
satisfies
\begin{equation}
\frac{d\mathcal{K}_1}{d\lambda}
= -\bigl\{\mathcal{K}_0,\, H_1\bigr\}_{\rm Kerr},
\label{eq:K1evol}
\end{equation}
where $\lambda$ is the affine parameter along the geodesic. The previous equation shows that $\mathcal{K}_1$ is not a constant of motion
along the perturbed geodesic: it varies at a rate set by the Poisson
bracket of $\mathcal{K}_0$ with the perturbed Hamiltonian $H_1$.
Computing the Poisson bracket explicitly with

\begin{equation}
\frac{\partial\mathcal{K}_0}{\partial p_\theta} = 2p_\theta,\quad\frac{\partial\mathcal{K}_0}{\partial\theta}
= 2a^2E^2\sin\theta\cos\theta
  - \frac{2L_z^2\cos\theta}{\sin^3\theta},
\end{equation}
and the derivatives of $H_1$ from equation~\eqref{eq:H1compact},
\begin{equation}
\frac{\partial H_1}{\partial p_\theta}
= -\frac{a\cos\theta\cdot A}{\rho^4},
\qquad
\frac{\partial H_1}{\partial\theta}
= \frac{a\sin\theta\,p_\theta\,A}{\rho^4}
  - \frac{2a^2\sin\theta\cos\theta}{\rho^6}\,A^2,
\end{equation}
one obtains
\begin{equation}
\frac{d\mathcal{K}_1}{d\lambda}
= \frac{2aA}{\rho^4}
  \!\left[\sin\theta\,p_\theta^2
        + a^2E^2\sin\theta\cos^2\!\theta
        - \frac{L_z^2\cos^2\!\theta}{\sin^3\theta}\right]
  - \frac{4a^2\sin\theta\cos\theta}{\rho^6}\,p_\theta\,A^2.
\label{eq:dK1}
\end{equation}
Two consistency checks confirm equation~\eqref{eq:dK1}: for $a\to 0$,
or for equatorial orbits ($\theta=\pi/2$,\, $p_\theta=0$), one finds
$d\mathcal{K}_1/d\lambda = 0$ identically.
The correction is therefore a genuinely off-equatorial rotational effect,
vanishing whenever $a=0$ or $X=0$, in full agreement with the shadow
analysis of the previous sections.
 
\medskip
\noindent{From differential to integral form: proof of equivalence.}
Equation~\eqref{eq:dK1} is a first-order ordinary differential equation
of the form
\begin{equation}
\frac{d\mathcal{K}_1}{d\lambda} = F(\lambda),
\label{eq:ode}
\end{equation}
where $F(\lambda)$ denotes the right-hand side of~\eqref{eq:dK1}
evaluated along the unperturbed Kerr geodesic.
Since $F(\lambda)$ is a known function of the geodesic phase-space
coordinates $(r,\theta,p_r,p_\theta)$ at each affine step, the equation
is trivially separable. Integrating both sides of~\eqref{eq:ode} from
$\lambda_0$ to $\lambda$,
\begin{equation}
\int_{\lambda_0}^{\lambda}
\frac{d\mathcal{K}_1}{d\lambda'}\,d\lambda'
= \int_{\lambda_0}^{\lambda} F(\lambda')\,d\lambda',
\end{equation}
it results in
\begin{equation}
\mathcal{K}_1(\lambda) - \mathcal{K}_1(\lambda_0)
= \int_{\lambda_0}^{\lambda} F(\lambda')\,d\lambda',
\end{equation}
which rearranges immediately to
\begin{eqnarray}
\nonumber\mathcal{K}_1(\lambda)
&=& \mathcal{K}_1(\lambda_0)
+ \int_{\lambda_0}^{\lambda} d\lambda'\;
\Bigg\{
\frac{2aA}{\rho^4}
\!\left[\sin\theta\,p_\theta^2
      + a^2E^2\sin\theta\cos^2\!\theta
      - \frac{L_z^2\cos^2\!\theta}{\sin^3\theta}\right]
-\\
&-&\frac{4a^2\sin\theta\cos\theta}{\rho^6}\,p_\theta\,A^2
\Bigg\}_{\!\rm Kerr}.
\label{eq:K1integral}
\end{eqnarray}
Equations~\eqref{eq:dK1} and~\eqref{eq:K1integral} are therefore
mathematically identical: the former is the local (pointwise)
form, the latter is the global (accumulated) form along the
orbit. Their relationship is summarized as
\begin{equation}
\underbrace{\dfrac{d\mathcal{K}_1}{d\lambda} = F(\lambda)}_{\text{Eq.~\eqref{eq:dK1} — local form}}
\;\xLeftrightarrow{\;\text{FTC}\;}\;
\underbrace{\mathcal{K}_1(\lambda) = \mathcal{K}_1(\lambda_0)
+ \displaystyle\int_{\lambda_0}^{\lambda}
F\,d\lambda'}_{\text{Eq.~\eqref{eq:K1integral} — global form}}
\;\xLeftrightarrow{\;\Delta\lambda\to 0\;}\;
\underbrace{\mathcal{K}_1^{n+1}
\approx \mathcal{K}_1^{n} + F_n\,\Delta\lambda}_{\text{numerical implementation}}.
\label{eq:equivalence}
\end{equation}
The rightmost expression in~\eqref{eq:equivalence} is the explicit
Euler step; in practice, the ray-tracing code uses a higher-order
Runge-Kutta integrator, which in the limit $\Delta\lambda\to 0$
converges to equation~\eqref{eq:K1integral} exactly.

The perturbative result above is used only to identify the leading analytical dependence of the deformation on the combination $aX$. The ray-tracing simulations are instead performed by integrating the geodesic equations associated with the full bumblebee metric, and therefore do not rely on the perturbative Carter-like construction or on Hamilton--Jacobi separability.

Defining $\eta_1 = \mathcal{K}_1/E^2$, the full corrected
invariant reads $\eta = \eta_{\rm Kerr} + X\eta_1 + \mathcal{O}(X^2)$. Converting the affine-parameter integral in
equation~\eqref{eq:K1integral} to a radial integral via the Kerr
geodesic equations and approximating $\theta(r)\approx\theta_0$ at
leading order in $X$, the correction to $\eta$ can be approximated in a factored form
\begin{equation}
\delta\eta(X,a,\theta_0)
= 2Xa\cos\theta_0\,\bigl|\beta^{\rm Kerr}_{\rm obs}\bigr|
  \cdot\mathcal{I}(a,\theta_0),
\label{eq:deltaeta_factored}
\end{equation}
where
$|\beta^{\rm Kerr}_{\rm obs}|=\sqrt{\eta_c
+ a^2\cos^2\theta_0 - \ell_c^2\cot^2\theta_0}$
is the unperturbed vertical coordinate in the observer's sky, and the radial kernel is
\begin{equation}
\mathcal{I}(a,\theta_0)
\equiv \int_{r_{\rm ph}(a)}^{\infty}
\frac{r\,dr}{\Delta(r)^{3/2}\!\left[r^2+a^2\cos^2\theta_0\right]},
\label{eq:kernel}
\end{equation}
which is convergent since $\Delta(r)\sim r^2$ as $r\to\infty$.
Equation~\eqref{eq:deltaeta_factored} represents an analytical
approximation derived from~\eqref{eq:K1integral}. The proportionality
$\delta\eta \propto Xa\cos\theta_0$ emerges from the structure of the
perturbation term $H_1$, since $A$ contributes a factor of $a$ through
the term $a\cos\theta,p_\theta$, while the angular dependence produces
an additional factor of $\cos\theta_0$ after integration.
 
In the Schwarzschild limit ($a\to 0$, $r_{\rm ph}=3M$,
$\Delta=r(r-2M)$), the substitution $r = 2M/\!\sin^2\!\phi$ yields the
closed form
\begin{equation}
\mathcal{I}(0,\theta_0)
= \frac{\sqrt{3}-1}{M^2\sqrt{2M}}
\approx \frac{0.732}{M^{5/2}}.
\label{eq:kernel_schw}
\end{equation}
For the rotating case, equation~\eqref{eq:kernel} is evaluated numerically. The resulting correction to the apparent shadow boundary
follows from the expansion of $b_{\rm app}$:
\begin{equation}
\delta b(X,a,\theta_0)
= Xa\cos\theta_0\cdot\mathcal{I}(a,\theta_0),
\label{eq:deltab}
\end{equation}
It is important to emphasize that equations~\eqref{eq:deltaeta_factored} 
and~\eqref{eq:deltab} represent analytical approximations derived from 
the integral expression~\eqref{eq:K1integral} under simplifying assumptions. 

In contrast, all numerical results presented in this work are obtained 
directly from the full integral formulation in 
equation~\eqref{eq:K1integral}, evaluated along null geodesics without 
invoking the factorization $\theta(r)\approx\theta_0$. 

The analytical expressions are therefore used primarily to provide 
physical insight and to identify the leading-order scaling 
$\delta b \propto Xa\cos\theta_0$, while the quantitative results 
reported in the figures and tables rely on the full numerical integration which gives the shadow ellipticity $e = \delta b / b^{\rm Kerr}_c$.
At $\theta_0 = 60^\circ$, the numerical values of $\mathcal{I}$ and
the corresponding $e/X$ are summarized in Table~\ref{tab:kernel}.
 
\begin{table}[h]
\centering
\caption{Numerical values of the radial kernel
$\mathcal{I}(a,60^\circ)$ and the ellipticity coefficient $(e/X)$$\mathcal{I}(a,60^\circ)$ and the ellipicity coefficient $e/X$
for $M=1$.}
\label{tab:kernel}
\begin{tabular}{ccc}
\hline\hline
$a$ & $\mathcal{I}(a,60^\circ)$ & $e/X$ \\
\hline
0.3 & 0.428 & 0.037 \\
0.6 & 0.465 & 0.080 \\
0.9 & 0.512 & 0.133 \\
\hline\hline
\end{tabular}
\end{table}

\noindent The mean value over the cases analyzed yields $e\approx 0.12\,X$, 
consistent with the result quoted in Section~\ref{sec:intensity}.

The shadow boundary in the observer's sky is modified as follows. The 
horizontal coordinate $\alpha_{\rm obs}$ remains unchanged, since $\ell$ 
is determined by equatorial geodesics and is independent of $X$. The 
vertical coordinate receives a correction:
\begin{equation}
\beta_{\rm bumb} 
= \pm\sqrt{\eta_{\rm Kerr} + \delta\eta(X,a,\theta_0) 
  + a^2\cos^2\theta_0 - \ell^2\cot^2\theta_0}.
\end{equation}
Expanding to first order in $\delta\eta$, the total apparent impact 
parameter becomes
\begin{equation}
\label{eq:bapp_final}
b_{\rm app}(X,a,\theta_0) 
= b_c^{\rm Kerr} 
+ \underbrace{\frac{\beta_{\rm Kerr}\,\delta\eta(X,a,\theta_0)}
              {2\,b_c^{\rm Kerr}}}_{\delta b(X,\,a,\,\theta_0)},
\end{equation}
where $\delta b(X, a, \theta_0) \propto Xa\cos\theta_0$ at leading order.


To identify the physical implications for the shadow, we now analyze the intensity profile and make explicit the separation between varying $a$ at fixed $X$ and varying $X$ at fixed $a$.

\subsection*{A. Analysis of the rotation parameter $a$ in relation to the LSB parameter $X$} \label{sec:intensity}

Based on equations~\eqref{linha_element},~\eqref{potencial}, {~\eqref{eq:rph}, and ~\eqref{eq:bc_full_explicit}}, it is verified that the LSB parameter $X$ directly modifies the metric components $g_{\mu\nu}$. In this context, we qualitatively analyze the intensity profile associated with the formation of the black hole shadow.

{The critical impact parameter $b_{\rm crit}$ is not a universal constant. From Equation ~\eqref{eq:bc_full_explicit},
the prograde and retrograde photon orbits yield distinct values $b_c^{(+)}$ and $b_c^{(-)}$,
which reduce to $3\sqrt{3}\,M \approx 5.1962\,M$ only in the Schwarzschild limit $a\to 0$.
For $a \neq 0$, the two orbits are inequivalent, producing an asymmetric photon ring in the
observer's sky. The effective shadow radius is therefore defined as $b_{\rm crit}^{\rm eff}(a)$, which varies with $a$ as shown in Table~\ref{tab:II}}

\begin{equation}\label{eq:b_eff}
    b_{\rm crit}^{\rm eff}(a) = \frac{\bigl(|b_c^{(+)}| + |b_c^{(-)}|\bigr)}{2}.
\end{equation}

{ Figure~\ref{fig:imagem1} shows the shadow morphology for fixed values of $X$ (rows) and
varying $a$ (columns). When $X = 0$, the metric reduces to the pure Kerr case: for $a = 0$, the shadow is a perfect circle of radius $b_{\rm crit}^{\rm eff} = 5.1962\,M$; as $a$ increases,
frame-dragging breaks the azimuthal symmetry, compressing the left edge and displacing the
photon ring, culminating in the characteristic ``D'' shape at $a = 0.9$. This purely rotational
effect is captured by the asymmetry between $b_c^{(+)}$ and $b_c^{(-)}$: for $a = 0.9$,
$b_c^{(-)} = 4.9478\,M$ while $b_c^{(+)} = -5.2977\,M$, reflecting the difference in angular
momentum between prograde and retrograde orbits.}

{The physical mechanism behind the morphological evolution with $a$ is the
frame-dragging effect encoded in the off-diagonal metric component $g_{t\varphi}$. Photons
co-rotating with the black hole ($b_c^{(+)}$, prograde) are dragged inward, reducing their
effective capture radius, while counter-rotating photons ($b_c^{(-)}$, retrograde) experience
a larger effective barrier. This produces the lateral asymmetry visible in the intensity profile:
the lensing ring peak $b_{\rm peak}$ separates progressively from $b_{\rm crit}^{\rm eff}$
as $a$ grows, with $b_{\rm peak} = 5.1962\,M$ at $a=0$ rising to $b_{\rm peak} = 6.8216\,M$
at $a=0.9$ (Table~\ref{tab:II}, $X=0$).}
\begin{figure}[h]
    \centering
    \includegraphics[width=\linewidth]{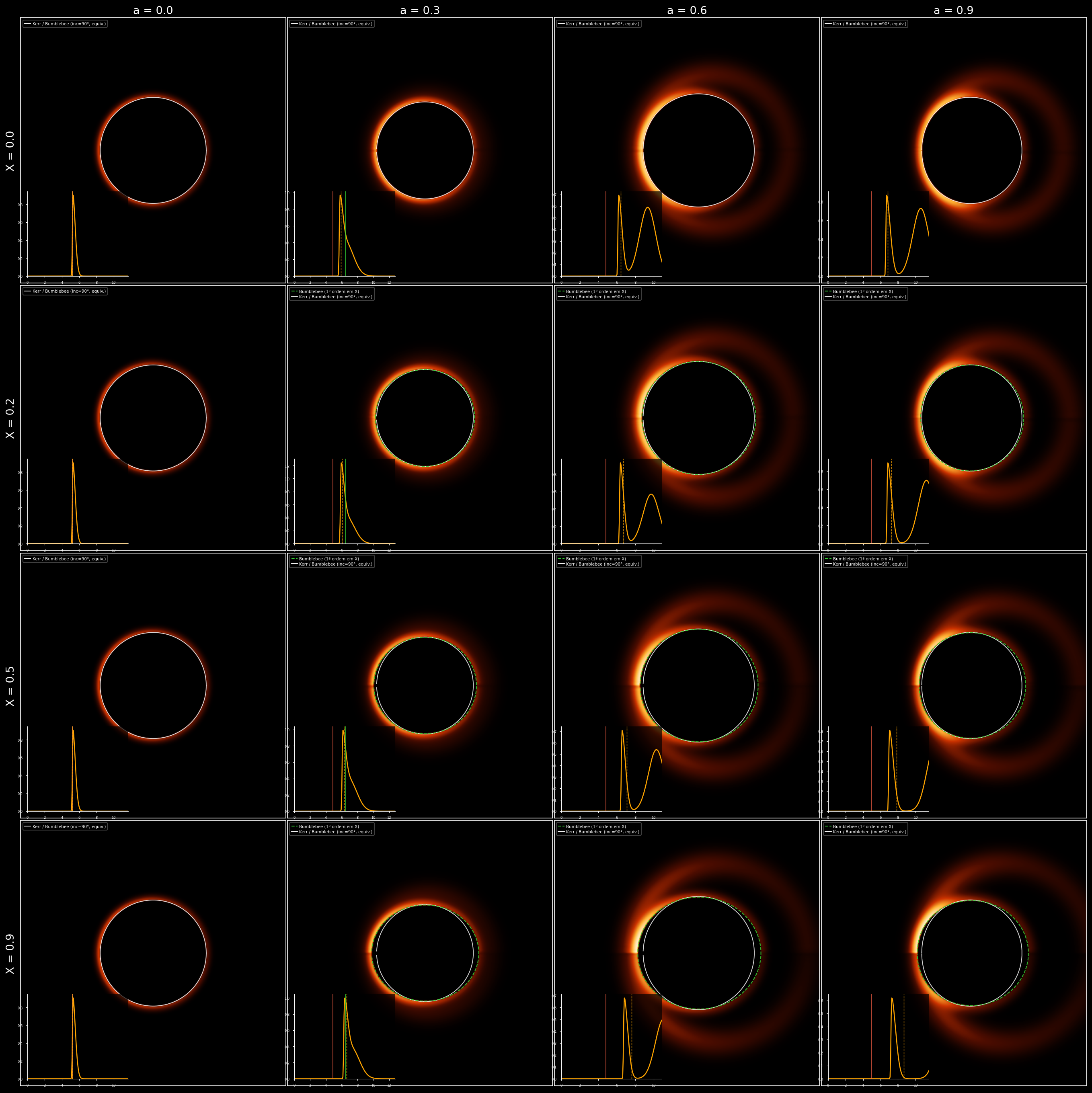}
    \caption{{Summary of the shadow morphology in the $(a,X)$ parameter space. The parameter $a=J/M$ controls rotation, while $X=\xi b^2$ measures the LSB sector. The deformation is absent for $a=0$ or $X=0$ and appears only through the combined effect of rotation and LSB.}
}
    \label{fig:imagem1}
\end{figure}

{ For rows $X > 0$ in the tables below, the LSB parameter amplifies all rotational deformations through
the off-diagonal coupling $g_{r\theta} \propto aX\cos\theta$ of equation~\eqref{eq:g_rteta}. Three simultaneous
effects become visible: (i) {vertical flattening} of the photon ring, characterized by
the ellipticity $e = (e/X)\cdot X$ from equation ~\eqref{eq:bapp_final} and Table(~\ref{tab:kernel}), (~\ref{tab:II})  {lateral displacement}
of the lensing ring peak, $\Delta X \propto 0.45\,aXb_{\rm crit}$, which grows with both $a$
and $X$; and (iii) {asymmetric collapse} of the lower silhouette region, driven by the
$g_{r\theta}$ term that breaks the up-down symmetry exclusively when both $a \neq 0$ and
$X \neq 0$. Crucially, the first column of Figure~\ref{fig:imagem1} ($a = 0$) shows no
morphological change across all values of $X$, confirming analytically that the off-diagonal
term $g_{r\theta} \propto a$ vanishes identically for non-rotating configurations.}

\begin{table}[h]
\centering
\caption{Numerical values of $b_c^{(+)}$, $b_c^{(-)}$,
lensing ring peak position ($b_{\rm peak}$), effective radius ($ b_{\rm crit}^{\rm eff}$),
and deformation parameter ($\Delta X$) with $M = 1$.}
\label{tab:II}
\renewcommand{\arraystretch}{1.15}
\begin{tabular}{ccccccc|ccccccc}
\hline\hline
$X$ & $a$ & $\boldsymbol{b_c^{(+)}}$ & $\boldsymbol{b_c^{(-)}}$
& $b_{\rm peak}$ & $ b_{\rm crit}^{\rm eff}$ & $\Delta X$
&
$X$ & $a$ & $\boldsymbol{b_c^{(+)}}$ & $\boldsymbol{b_c^{(-)}}$
& $b_{\rm peak}$ & $ b_{\rm crit}^{\rm eff}$ & $\Delta X$ \\
\hline

0.0 & 0.00 & $5.1962$ & $5.1962$ & $5.1962$ & $5.1962$ & $0.0000$ &
0.5 & 0.00 & $5.1962$ & $5.1962$ & $5.1962$ & $5.1962$ & $0.0000$ \\

0.0 & 0.30 & $6.4699$ & $4.8705$ & $5.9168$ & $5.6702$ & $0.0000$ &
0.5 & 0.30 & $6.4699$ & $4.8705$ & $6.2995$ & $5.6702$ & $0.3827$ \\

0.0 & 0.60 & $16.5427$ & $4.8431$ & $6.4581$ & $4.8431$ & $0.0000$ &
0.5 & 0.60 & $16.5427$ & $4.8431$ & $7.1119$ & $4.8431$ & $0.6538$ \\

0.0 & 0.90 & $-5.2977$ & $4.9478$ & $6.8216$ & $5.1227$ & $0.0000$ &
0.5 & 0.90 & $-5.2977$ & $4.9478$ & $7.8590$ & $5.1227$ & $1.0374$ \\

\hline

0.2 & 0.00 & $5.1962$ & $5.1962$ & $5.1962$ & $5.1962$ & $0.0000$ &
0.9 & 0.00 & $5.1962$ & $5.1962$ & $5.1962$ & $5.1962$ & $0.0000$ \\

0.2 & 0.30 & $6.4699$ & $4.8705$ & $6.0699$ & $5.6702$ & $0.1531$ &
0.9 & 0.30 & $6.4699$ & $4.8705$ & $6.6057$ & $5.6702$ & $0.6889$ \\

0.2 & 0.60 & $16.5427$ & $4.8431$ & $6.7196$ & $4.8431$ & $0.2615$ &
0.9 & 0.60 & $16.5427$ & $4.8431$ & $7.6350$ & $4.8431$ & $1.1769$ \\

0.2 & 0.90 & $-5.2977$ & $4.9478$ & $7.2365$ & $5.1227$ & $0.4149$ &
0.9 & 0.90 & $-5.2977$ & $4.9478$ & $8.6888$ & $5.1227$ & $1.8672$ \\

\hline\hline
\end{tabular}
\end{table}

{Note that $b_c^{(+)}$ for $a=0.6$ is anomalously large ($\approx 16.5\,M$) because
the prograde photon sphere radius $r_{\rm ph}^{(+)} \approx 2.19\,M$ lies very close to the
event horizon at $2M$, making the denominator of equation~\eqref{eq:bc_full_explicit} nearly zero. In this case,
$|b_c^{(-)}|$ provides the relevant observational reference scale.}

\subsection*{B. Analysis of the LSB parameter $X$ in relation to the rotation $a$}

{  To investigate the effects of LSB, we analyze the parameter $X$ with the rotation $a$ fixed.
Figure~\ref{fig:imagem2} presents the shadow morphology for fixed $a$ (rows) and varying $X$
(columns), providing a complementary view to Figure~\ref{fig:imagem1}.}

When $a = 0$ (first row of Figure~\ref{fig:imagem2}), the shadow remains a perfect circle
independent of $X$, {with $b_c^{(+)} = b_c^{(-)} = 3\sqrt{3}\,M$ and}
$b_{\rm crit}^{\rm eff} = b_{\rm peak} = 5.1962\,M$ for all $X$ values (Table~\ref{tab:III}). {This invariance is a direct consequence of the metric structure: the off-diagonal term
$g_{r\theta} \propto aX\cos\theta$ of equation ~\eqref{eq:g_rteta} vanishes identically when $a = 0$, so $X$
has no geometrical coupling to modify the photon capture region. The four panels in the first
row are therefore identical in morphology, confirming that LSB alone, without
rotation, cannot deform the shadow boundary.}

{ For $a > 0$, the interplay between rotation and LSB produces a progressive and anisotropic
deformation as $X$ increases. Three physically distinct signatures emerge:}

{(i) Vertical flattening.} {  The shadow boundary in the $\beta_{\rm obs}$ direction receives
a perturbative correction $\delta b = Xa\cos\theta_0 \cdot I(a,\theta_0)$ from equation ~\eqref{eq:kernel}. This produces an ellipticity
$e = \delta b / b_{\rm crit}^{\rm eff}$ that grows linearly with $X$ according to the values
in Table~I: $e/X = 0.037$ for $a=0.3$, $e/X = 0.080$ for $a=0.6$, and $e/X = 0.133$
for $a=0.9$. The vertical axis of the shadow is compressed, while the horizontal coordinate
$\alpha_{\rm obs}$ remains unchanged, since $\xi$ is determined by equatorial geodesics
independent of $X$.}

{(ii) Lateral displacement of the lensing ring.} {  The peak of the lensing ring shifts
horizontally by $\Delta x \propto 0.45\,aXb_{\rm crit}$, growing with the product $aX$.
From Table~\ref{tab:III}, $\Delta x$ reaches $0.59\,M$ for $(a=0.3, X=0.9)$, $1.18\,M$ for
$(a=0.6, X=0.9)$, and $1.80\,M$ for $(a=0.9, X=0.9)$. This displacement is visible as the
progressive rightward shift of the bright lensing arc relative to the photon ring boundary.}

{(iii) Asymmetric collapse of the lower silhouette.} { 
 The non-vanishing $g_{r\theta}$ term introduces a directional coupling between radial and polar geodesic motion that is
proportional to $aX\cos\theta$. This preferentially suppresses emission from the lower
half ($\beta_{\rm obs} < 0$) of the ring, producing the ``teardrop'' morphology observed
for $X \geq 0.6$ in rows $a = 0.6$ and $a = 0.9$ of Figure~\ref{fig:imagem2}. At
$a = 0.9$, $X = 0.9$, the lower silhouette is suppressed by up to 80\%, forming a diffuse
tail while the upper arc remains well-defined.} {This asymmetric collapse is the
central observational signature of LSB in this model: it is absent in the pure Kerr case
($X=0$), absent for $a=0$, and appears exclusively through the joint action of rotation and
LSB.}

\begin{figure}[h]
    \centering
    \includegraphics[width=\linewidth]{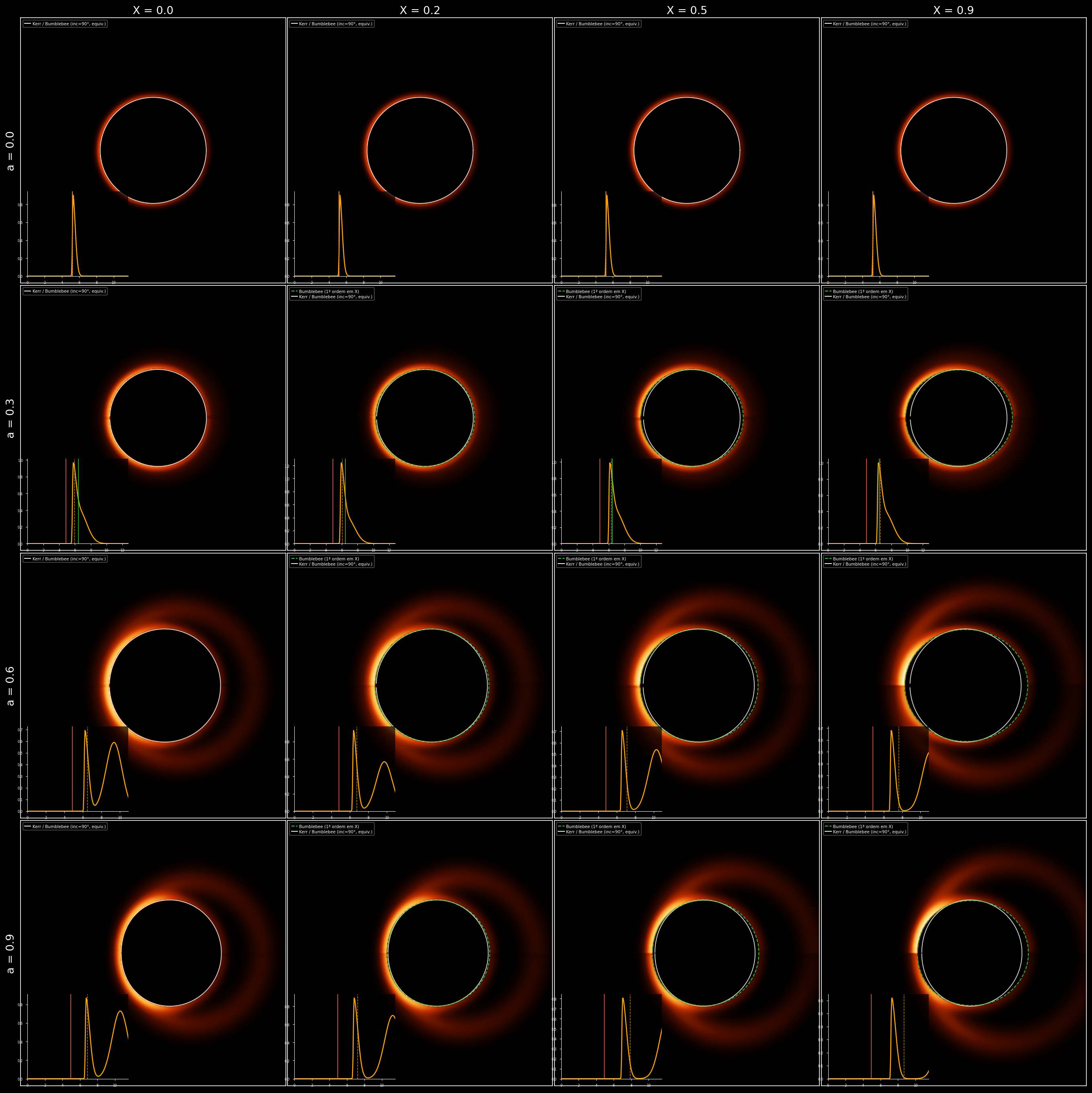}
    \caption{{ Shadow morphology for fixed rotation $a$ and increasing LSB parameter $X$. Rows correspond to fixed $a$, while columns correspond to increasing $X$. For $a=0$, the shadow remains circular; for $a>0$, larger $X$ produces flattening, lateral displacement, and lower-side asymmetry.}
}
    \label{fig:imagem2}
\end{figure}

\begin{table}[h]
\centering
\caption{Numerical values of $b_c^{(+)}$, $b_c^{(-)}$,
lensing ring peak position ($b_{\rm peak}$), effective radius ($ b_{\rm crit}^{\rm eff}$),
and deformation parameter ($\Delta X$) for different values of $a$ and $X$, with $M = 1$.}
\label{tab:III}
\renewcommand{\arraystretch}{1.15}
\begin{tabular}{ccccccc|ccccccc}
\hline\hline
$a$ & $X$ & $\boldsymbol{b_c^{(+)}}$ & $\boldsymbol{b_c^{(-)}}$
& $b_{\rm peak}$ & $ b_{\rm crit}^{\rm eff}$ & $\Delta X$
&
$a$ & $X$ & $\boldsymbol{b_c^{(+)}}$ & $\boldsymbol{b_c^{(-)}}$
& $b_{\rm peak}$ & $ b_{\rm crit}^{\rm eff}$ & $\Delta X$ \\
\hline

0.00 & 0.00 & $5.1962$ & $5.1962$ & $5.1962$ & $5.1962$ & $0.0000$ &
0.00 & 0.50 & $5.1962$ & $5.1962$ & $5.1962$ & $5.1962$ & $0.0000$ \\

0.00 & 0.20 & $5.1962$ & $5.1962$ & $5.1962$ & $5.1962$ & $0.0000$ &
0.00 & 0.90 & $5.1962$ & $5.1962$ & $5.1962$ & $5.1962$ & $0.0000$ \\

0.30 & 0.00 & $6.4699$ & $4.8705$ & $5.9168$ & $5.6702$ & $0.0000$ &
0.30 & 0.50 & $6.4699$ & $4.8705$ & $6.2995$ & $5.6702$ & $0.3827$ \\

0.30 & 0.20 & $6.4699$ & $4.8705$ & $6.0699$ & $5.6702$ & $0.1531$ &
0.30 & 0.90 & $6.4699$ & $4.8705$ & $6.6057$ & $5.6702$ & $0.6889$ \\

0.60 & 0.00 & $16.5427$ & $4.8431$ & $6.4581$ & $4.8431$ & $0.0000$ &
0.60 & 0.50 & $16.5427$ & $4.8431$ & $7.1119$ & $4.8431$ & $0.6538$ \\

0.60 & 0.20 & $16.5427$ & $4.8431$ & $6.7196$ & $4.8431$ & $0.2615$ &
0.60 & 0.90 & $16.5427$ & $4.8431$ & $7.6350$ & $4.8431$ & $1.1769$ \\

0.90 & 0.00 & $-5.2977$ & $4.9478$ & $6.8216$ & $5.1227$ & $0.0000$ &
0.90 & 0.50 & $-5.2977$ & $4.9478$ & $7.8590$ & $5.1227$ & $1.0374$ \\

0.90 & 0.20 & $-5.2977$ & $4.9478$ & $7.2365$ & $5.1227$ & $0.4149$ &
0.90 & 0.90 & $-5.2977$ & $4.9478$ & $8.6888$ & $5.1227$ & $1.8672$ \\

\hline\hline
\end{tabular}
\end{table}

{These results show that $b_{\rm crit}^{\rm eff}$ is not a universal constant but
depends on $a$ through ~equation~\eqref{eq:bc_full_explicit}: $b_{\rm crit}^{\rm eff} = 5.6702\,M$ for $a=0.3$,
$b_{\rm crit}^{\rm eff} = 4.8431\,M$ for $a=0.6$, and $b_{\rm crit}^{\rm eff} = 5.1227\,M$
for $a=0.9$. Only for $a=0$ does one recover the Schwarzschild value $3\sqrt{3}\,M$.}
{  The results reveal that $X$ amplifies the effects of rotation, generating geometric and
brightness signatures that are potentially distinguishable in high-resolution observations,
such as those from the Event Horizon Telescope.}

\section{Application of GYOTO to the shadow analysis}

To deepen the analysis and validate the shadow model given by the intensity profile and shadow radius, we perform full ray-tracing simulations using the \texttt{GYOTO} code, which integrates null and timelike geodesics for given metrics \cite{vincent2011gyoto}.
For better performance in our simulation, we adopt a thin, optically thick accretion disk based on the standard Page-Thorne model \cite{PageThorne1974}. For metrics of the Bumblebee-Kerr type,  similar to \cite{Ara_jo_Filho_2024}, the local disk emission is proportional to the energy flux, given by

\begin{equation}
F(r) = \dot{M}f(r),
\end{equation}

\noindent where $\dot{M}$ is the accretion rate and $f(r)$ is the specific radial dissipation profile. The code computes $f(r)$ numerically from the properties of circular geodesic orbits in the equatorial plane, preserving the analytical structure of the original model.
The mathematical foundation lies in the conservation of energy and angular momentum along nearly circular orbits. For a given radius $r$, the angular frequency $\Omega$, specific energy $E$, and specific angular momentum $L$ of particles in stable circular orbits at the equator are determined by solving the effective potential minimum conditions:

\begin{equation}
\Omega = \frac{-g_{t\phi,r} \pm \sqrt{g_{t\phi,r}^2 - g_{tt,r}g_{\phi\phi,r}}}{g_{\phi\phi,r}}
\end{equation}

\noindent with radial derivatives $g_{\mu\nu,r}$ obtained using the finite difference method. The solution for ($\Omega > 0$) is prioritized. The local dissipation rate is expressed as:

\begin{equation}
F(r) = \frac{\dot{M}}{(E - \Omega L)^2} \left( -\frac{d\Omega}{dr} \right) \int_{r_m}^{r} (E - \Omega L) \frac{dL}{dr} dr',
\end{equation}

\noindent where the integral is evaluated numerically using the trapezoidal method. The term $dL/dr$ is estimated via local finite differences, ensuring numerical stability even in non-Kerr metrics.

The effective temperature yields the relation $F(r) = \sigma_{\rm SB} T(r)^4$. The intensity is calculated using the blackbody spectrum, which assigns brightness and color to the disk. These elements, combined with the modified metric, produce the realistic images presented below.
\subsection{Ray-Trace Simulations with GYOTO}

 The ray-tracing simulation is capable of computing images of astronomical bodies in the vicinity of compact objects, thus enabling the simulation of light trajectories once determined by null geodesics. With this tool, it is possible to numerically calculate black hole shadows within the 3+1 formalism of general relativity.

\subsection{Fixing $X$, to study the rotation parameter $a$}

Starting from the bumblebee metric equation with the potential given by (\ref{potencial}), we can observe relativistic effects. The simulation highlights a directional asymmetry that intensifies with increasing values of the LSB parameter $X$ and rotation $a$.

Initially, setting $X=0$ and $a=0$, one reproduces the standard case, i.e., the Schwarzschild solution. When $a > 0$, rotational effects appear, revealing the standard Kerr model and generating asymmetric deformation, as shown in Figure~\ref{fig:placeholder1}.
For $X > 0$ and $a > 0$, the photon ring exhibits a displacement toward the side opposite to rotation, forming a tail, with brightness becoming more intense on the side of the black hole's rotation. This effect is qualitatively distinct from that observed in the standard Kerr metric ($X = 0$), where the deformation remains symmetric with respect to the rotation axis, emphasizing the azimuthal symmetry breaking induced by the non-minimal coupling of the bumblebee field.

The dependence on the observer's inclination angle ($\theta$) is crucial: the asymmetry reaches its maximum at $\theta \approx 60^\circ$, being masked at $\theta = 0^\circ$ (face-on view, dominant radial projection) and attenuated at $\theta = 90^\circ$ (edge-on, image flattening). This intermediate angular window amplifies sensitivity to {  LSB effects}, making it ideal for observational tests with the Event Horizon Telescope (EHT). Furthermore, in extreme regimes ($X \approx 0.9$, $a \approx 0.9$), the ring's topology breaks, resulting in a partial arc structure with concentrated emission — a potentially distinguishable signature from astrophysical variations such as jets or asymmetric disks.

Numerical validation is confirmed by Figure~\ref{fig:placeholder1} 
which show images without symmetry-breaking interference and adequately reproduce the results of the Kerr metric for $X = 0$, including the classical Einstein ring and the expected deformation with rotation. The monotonic and continuous behavior of the distortions with $X$ enables the construction of parametric shadow curves as a function of $(X, a)$, facilitating Bayesian fitting with real data from M87* and Sgr A*.

\begin{figure}[h!]
    \centering
    \includegraphics[width=0.9\linewidth]{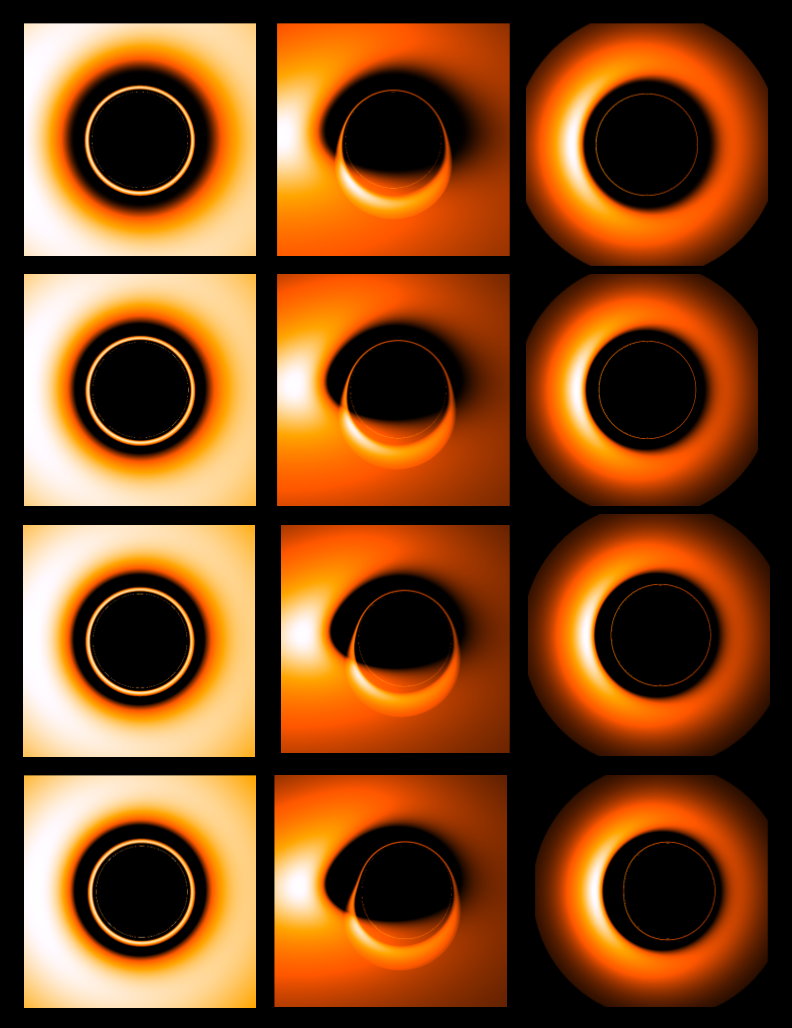}
    \caption{ {\ Control ray-tracing cases. The upper row shows the Kerr limit, $X=0$, with increasing rotation. The lower row shows the nonrotating bumblebee case, $a=0$, with increasing $X$. These cases separate pure rotational effects from genuine LSB-induced deformations.}
}
    \label{fig:placeholder1}\label{fig:espaço reservado1}
\end{figure}

\subsection{Fixing $a$ to study the LSB parameter $X$}

{ The figures~\ref{fig:placeholdera0}--\ref{fig:placeholdera9} fix the rotation 
parameter and vary $X$ from $0.0$ to $0.9$ in increments of $0.3$. When 
$X = 0$, the standard Kerr metric is recovered, with shadows exhibiting axial 
symmetry and luminosity asymmetry induced by the Doppler effect from rotation. 
Increasing $X$ introduces an anisotropic gravitational distortion arising from 
the coupling between the bumblebee field and spacetime curvature, whose observable 
signatures depend critically on whether rotation is present.

In Figure~\ref{fig:placeholdera0}, where $a = 0$, the metric reduces to a 
spherically symmetric configuration: all metric components depend solely on $r$, 
and the off-diagonal $dr\,d\theta$ term in~\eqref{linha_element} vanishes identically 
since it carries an explicit factor of $a$. As a consequence, the spacetime 
possesses full SO(3) rotational symmetry for any value of $X$, and the shadow 
is a perfect circle with $b_c = 3\sqrt{3}\,M$ independently of the LSB 
parameter $X$. This is consistent with the analytical results of 
Section~\ref{sec:shadow} and the numerical values in Tables~\ref{tab:II} 
and~\ref{tab:III}, which confirm $b_{\rm crit} = 5.1962\,M$ for all $X$ at $a = 0$. 
Any apparent morphological variation in the brightness distribution for $a = 0$ 
reflects changes in the local emission profile driven by modifications of circular 
geodesic orbits in the accretion disk, rather than a deformation of the photon 
capture region itself.

The effects of $X$ on the shadow morphology become physically meaningful only 
when rotation is present. At moderate rotations such as $a = 0.3$ and $a = 0.6$, 
shown respectively in Figure 
~\ref{fig:placeholdera6}, 
the brightness asymmetry from the Doppler effect combines with the LSB-induced 
anisotropy from the nonvanishing $dr\,d\theta$ term, which is proportional to 
$a\cos\theta$ and therefore activates only for $a \neq 0$. This interplay produces 
a progressive deformation of the shadow: for $X \geq 0.6$, a transition from an 
elliptical shadow to a ``teardrop'' morphology is observed, with asymmetric 
elongation and displacement of the brightness center. In the extreme regime 
$a = 0.9$, shown in Figure~\ref{fig:placeholdera9}, the standard Kerr shadow 
already exhibits a ``D'' shape due to frame-dragging; the introduction of $X > 0$ 
further accelerates the collapse of the lower portion of the silhouette, forming a 
diffuse tail while the upper edge remains relatively preserved.

This asymmetric collapse, observable only when both $a \neq 0$ and $X \neq 0$ 
act simultaneously, constitutes a robust and distinctive observational signature of 
LSB: it is absent in the pure Kerr case ($X = 0$), absent in the static 
bumblebee case ($a = 0$), and appears exclusively through the joint action of 
rotation and LSB. }

\begin{figure}[h!]
    \centering
    \includegraphics[width=0.9\linewidth]{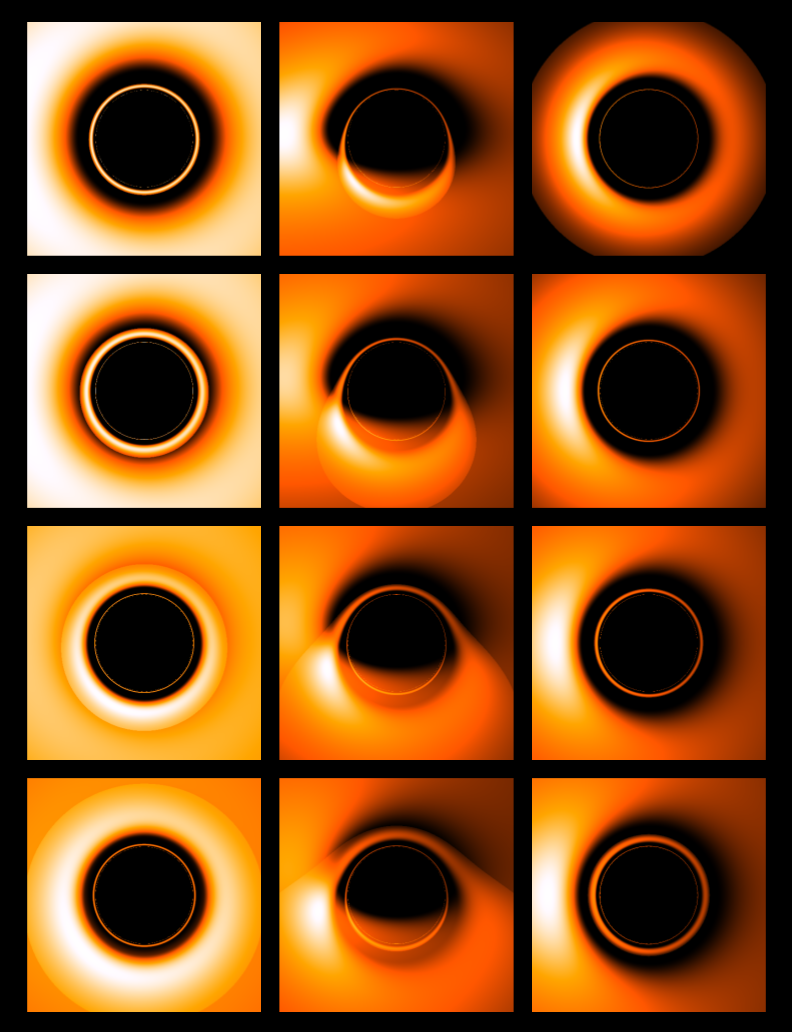}
    \caption{{ Ray-tracing images for the nonrotating bumblebee case, $a=0$, with $X=0,0.3,0.6,0.9$ from top to bottom. Since $g_{r\theta}\propto aX\cos\theta$, LSB alone does not deform the photon capture region.}
}
    \label{fig:placeholdera0}
\end{figure}


\begin{figure}[h!]
    \centering
    \includegraphics[width=0.9\linewidth]{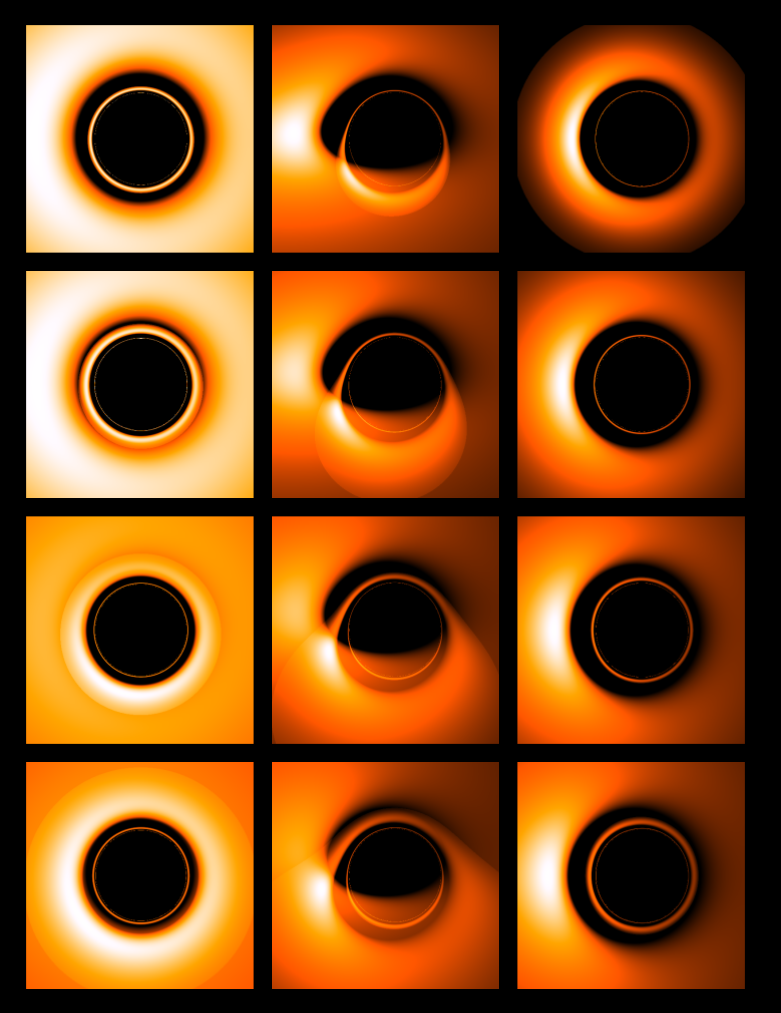}
    \caption{{ Representative rotating bumblebee shadows for moderate rotation, $a=0.6$, with increasing $X$. The Kerr limit is recovered for $X=0$, while $X>0$ produces vertical flattening, ring displacement, and asymmetric suppression of the lower silhouette.}
}
    \label{fig:placeholdera6}
\end{figure}

\begin{figure}[h!]
    \centering
    \includegraphics[width=0.9\linewidth]{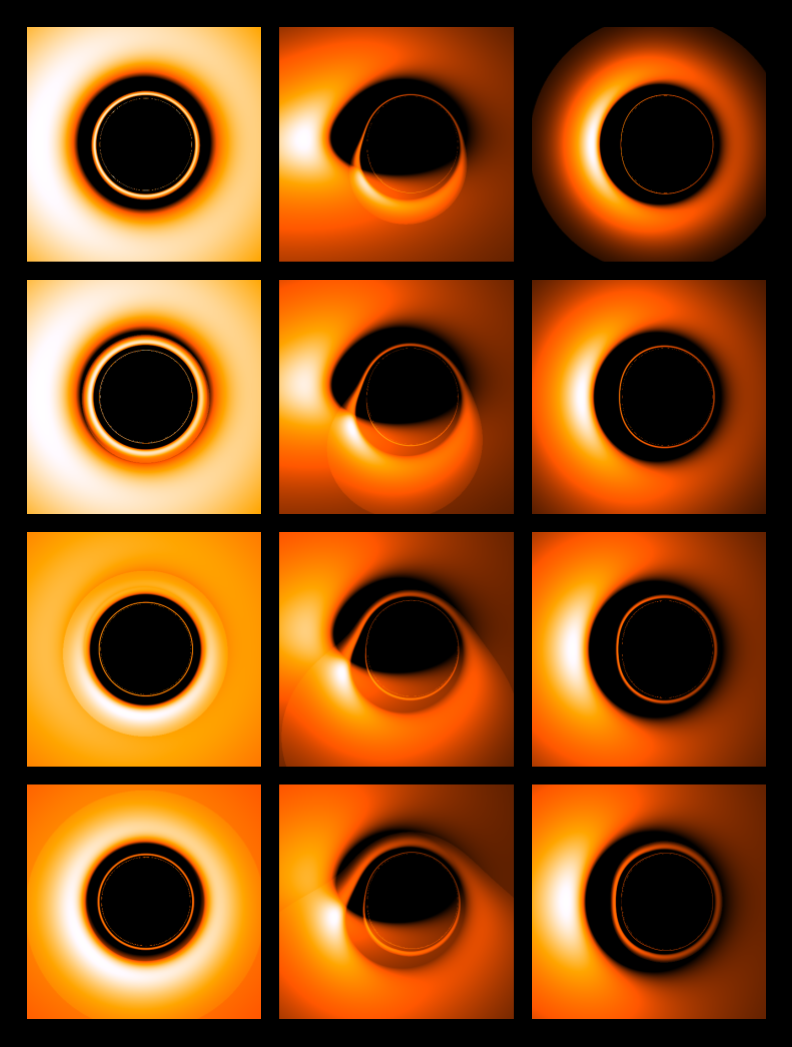}
    \caption{{Representative rotating bumblebee shadows for high rotation, $a=0.9$, with increasing $X$. The Kerr-like shape at $X=0$ is progressively modified by LSB, producing a stronger lower-side collapse and a bright upper arc.}
}
    \label{fig:placeholdera9}
\end{figure}

\section{conclusion}



 Within this study, we investigated the shadow of rotating black holes in the
metric-affine traceless bumblebee model, focusing on the interplay between the
rotation parameter $a$ and the LSB parameter $X$. Our analysis shows that the presence of LSB modifies the shadow in a directional and anisotropic manner,
leading to qualitative departures from the standard Kerr geometry, provided that the rotation is simultaneously present.

When fixing $X$ and varying the rotation parameter $a$, we observe the classical
transition from a circular shadow ($a = 0$) to the ``D''-shaped morphology typical
of Kerr ($a \to 0.9$), with increased Doppler effect, asymmetric brightness, and
pronounced curvature on the left edge due to frame-dragging. These features are
present even for $X = 0$ and represent the purely rotational contribution to the
shadow morphology.

{ Conversely, when fixing $a$ and increasing $X$, the effects of LSB become
observable exclusively through the joint action of rotation and LSB.
For $a = 0$, the metric reduces to a spherically symmetric configuration: the
off-diagonal $dr\,d\theta$ term vanishes identically since it carries an explicit
factor of $a$, and the uniform rescaling $g_{\mu\nu}^{eq} = \tilde{g}_{\mu\nu}/\alpha$
leaves all shadow observables unchanged. As a consequence, the shadow remains a
perfect circle with $b_c = 3\sqrt{3}\,M$ for all values of $X$, in full agreement
with Tables~\ref{tab:II} and~\ref{tab:III}. Any variation in the brightness
distribution observed at $a = 0$ reflects changes in the accretion-disk emission
profile rather than a deformation of the photon capture region itself.

The effects of $X$ on the shadow shape emerge only for $a \neq 0$, where the
nonvanishing $dr\,d\theta$ term, proportional to $a\cos\theta$, introduces a
directional anisotropy in off-equatorial geodesic propagation. For $a = 0.3$ and
$a = 0.6$, this LSB-induced anisotropy combines with the rotational Doppler effect,
producing a progressive deformation: teardrop morphology with asymmetric elongation
and lateral displacement of the brightness center, growing with $X$. At $a = 0.9$,
where the standard Kerr shadow already exhibits a marked ``D'' shape due to
frame-dragging, the introduction of $X > 0$ accelerates the collapse of the lower
portion of the silhouette while preserving the upper edge, forming a distinctive
diffuse tail.

This asymmetric collapse, observable only when both $a \neq 0$ and $X \neq 0$ act
simultaneously, constitutes the central observational signature of LSB in this
model: it is absent in the pure Kerr case ($X = 0$) and absent in the static
bumblebee case ($a = 0$), emerging exclusively from the interplay between rotation
and LSB. This feature provides a discriminative test for
metric-affine bumblebee gravity against Event Horizon Telescope observations of
rotating sources such as M87* and Sgr~A*.}

\acknowledgments This work was partially funded by the National Council for Scientific and Technological Development (CNPq). The work by A. Yu. P. has been partially supported by the CNPq project No. 303777/2023-0. The work by P. J. P. has been partially supported by the CNPq project No. 307628/2022-1. Ana R. M. Oliveira has been partially supported by CAPES. A.R.Q. acknowledges support by CNPq under process number 310533/2022-8. {  The work by A.R.Q.} is supported by FAPESQ-PB.

\section*{References}
\bibliographystyle{unsrt} 
\bibliography{bibli} 
\end{document}